\documentstyle[12pt,psfig,epsf]{article}
\newcommand{\be}{\begin{equation}}
\newcommand{\qee}{\end{equation}}
\newcommand{\bea}{\begin{eqnarray}}
\newcommand{\eea}{\end{eqnarray}}

\begin{document}

\hfill  NTZ 19/2000

\hfill  NRCPS-HE-2000-20

\vspace{3cm}

{\begin{center}
{\LARGE {Electromagnetic scattering on 
D3-brane spikes\footnote{Dedicated to the memory of Professor 
Gurgen Sahakian\\ http://server.physdep.r.am/~gsahak/ }} } \\ [12mm]
{\large  Roland Kirschner$^{a}$ and
George K. Savvidy$^{a,b}$ \\[3mm] } 
\end{center} 
} 

\vspace{0.5cm}

{\begin{center}
\item[$^a$] Naturwissenschaftlich-Theoretisches Zentrum \\
Institut f\"{u}r Theoretische Physik, Universit\"{a}t Leipzig, \\
Augustusplatz 10, D-04109 Leipzig, Germany
\item[$^b$] 
National Research Center Demokritos,\\
Ag. Paraskevi, GR-15310 Athens, Hellenic Republic  
\end{center} 
} 

\vspace{2cm} 
\noindent

\centerline{{\bf Abstract}}

\vspace{23pt}
\noindent

We consider scattering of  electromagnetic plane waves on a D3-brane 
spike which emanates normal to D3-barne in the extra space direction.
We are interested in studying physical effects on D3-brane which are 
produced by a spike attached to D3-brane. We have observed that the 
spike sucks almost all electromagnetic radiation and therefore acts 
like a black hole. This is
because absorption cross section for j=1 tends to a  constant at low energy 
limit. This behaviour is appealing for a string interpretation of the 
spike soliton because the propagation of $j=1$ mode is indeed distinctive. 
Instead, the scattered part of the radiation on a D3-brane 
tends to zero demonstrating non-Thompson behaviour. 


\newpage

\pagestyle{plain}

{\it Introduction.} The Dp-brane is a dynamical extended 
object which can fluctuate in 
shape and position and these fluctuations are described 
by open string massless modes \cite{rr,pol,leigh}. 
For a single Dp-brane they are massless vector and  spinor states
of ten-dimensional $N=1$ supersymmetric $U(1)$ gauge theory. The 
massless field $A_{\mu}(x^{\nu}), \mu,\nu = 0,...,p$ 
propagates as gauge bosons on the p-brane worldvolume, 
while the other components of the vector potential 
$A_{p+1}(x^{\nu}),...,A_9(x^{\nu})$ describe
the transverse deviations of the Dp-brane 
$x_{p+1}\equiv \phi_{p+1},...,x_9\equiv \phi_{9}$.. The low energy
effective action for these fields is the Dirac-Born-Infeld 
action \cite{leigh,sch}.

Callan and Maldacena \cite{cm,gibbons,howe} showed that the Dirac-Born-Infeld action
supports solitonic configurations describing F(fundamental)-strings 
and D(Dirichlet)-strings  
growing out of the original D3-brane. These configurations have nonzero 
world-volume gauge field and transverse scalar field excited. The 
gauge field describes a point electric or magnetic 
charge arising from the end-point of the 
attached string and the scalar field represents a deformation of the 
D3-brane in the form of infinite spike (see Figure 1). These
spike configurations are protruding from D3-brane universe into
extra space directions and it is interesting to study  
physical effects which they produce on D3-brane. 
In particular it is interesting to consider electromagnetic scattering 
on D3-brane spike. 

The spike soliton, F-string, which satisfies 1/2-BPS conditions 
of the $D=4$ $N=4$  supersymmetric electrodynamics is
\cite{cm},     
\be                                                                         
\vec{E} = \frac{e^2}{r^2}\vec{e_r}~,~~~~\vec{\partial}x_9                     
= \frac{e^2}{r^2}\vec{e_r}~,~~~~x_9 = -\frac{e^2}{r}~,                          
\label{spike}                                                               
\qee                                                                         
where $2e^2 = g_s$ is a unit charge.                                       
Here the scalar field represents a geometrical spike,                     
and the electric field insures that the string carries uniform NS           
charge along it. It was demonstrated            
in \cite{cm} that the infinite electrostatic energy of the point       
charge can be reinterpreted as being due to the infinite length of the      
attached string. The energy per unit length comes from the electric field   
and corresponds exactly to the fundamental string tension. If one replace 
$e^2$ by $Ne^2$ in the above solution, then it will describe $N$ 
superposed strings ending on a single D3-brane. Strict validity 
of the Born-Infeld action requires that derivatives of the 
field strengths, measured in units of $\alpha^{'}$, should be small,
that is when $Ne^2 \gg 1$.  In addition to ignore 
gravitational effects one should also require $e^4 N \ll 1$, 
that is $e^2 \gg 1/N \gg e^4$ \cite{cm}. 

As it was
shown in \cite{cm,larus,larus1,rey} small fluctuations which are normal to both the 
string and the brane are mostly reflected back with a 
$phase~ shift  \rightarrow \pi$ thus realizing dynamically Polchinski's
open string Dirichlet boundary condition.
In \cite{konstantin,thesis,bak,grandi} it was also demonstrated that P-wave 
excitations (j=1) which are coming down the string with a
polarization along a direction parallel to the brane are almost completely
reflected just as in the case of all-normal excitations, but now the end of
the string moves freely on the 3-brane. This corresponds to 
the reflection of the geometrical 
fluctuation with a $phase~ shift  \rightarrow 0$, thus realizing 
dynamically Polchinski's open string Neumann boundary condition.
The results obtained seems to have a larger regime of validity and can
partly be understood in light of the fact that BPS spike is a solution to the full 
equations of motion following from string theory \cite{larus11}.

\begin{figure}
\centerline{\hbox{\psfig{figure=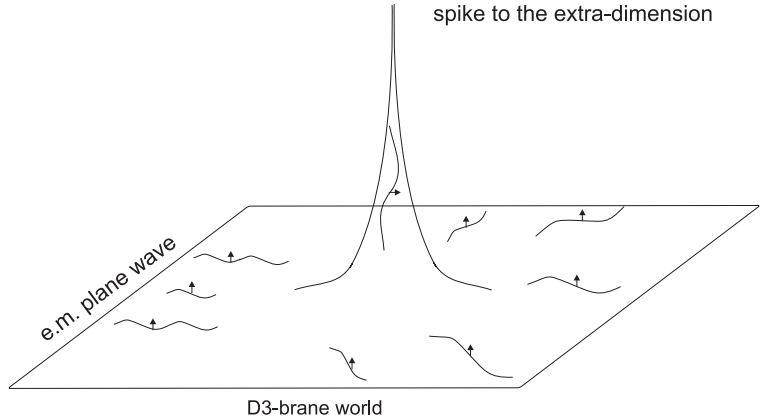,width=10cm}}}
\caption[fig]{This configuration describes D3-brane spike 
growing out of the original D3-brane and can be interpreted as a string 
attached to a D3-brane. We consider scattering of  electromagnetic 
plane waves on a D3-brane 
spike which emanate normal to D3-barne in the extra space direction. 
We have observed that the spike sucks almost all electromagnetic radiation, 
because absorption cross section tends to a constant at low energy limit,
instead, the scattered part of radiation on a D3-brane 
tends to zero demonstrating non-Thompson behaviour,
universal for the charged objects in electrodynamics.
}
\label{fig}
\end{figure}

In \cite{konstantin} it was also observed 
the electromagnetic dipole radiation which escapes to infinity
from the place where the string is attached to the D3-brane. 
It was shown that in the low energy limit
$R \omega \rightarrow 0$ ($R^2 = e^2 (2\pi \alpha^{'})$), 
i.e. for wavelengths much larger than the string
scale   a small fraction $ \sim \omega^4$ 
of the energy escapes to infinity in
the form of electromagnetic dipole radiation (j=1). 
The physical interpretation was that a string attached to the D3-brane
manifests itself as an electric charge, and waves on the string cause
the end point of the string to freely oscillate and produce
electromagnetic dipole radiation in the asymptotic outer region of the D3-brane.
Thus dynamics of the spike,
as probed through small fluctuations, agree with existing string behaviour.

The propagation of higher  angular momentum modes has been analyzed in   
\cite{kastor}, where authors came to the conclusion that they also 
propagate  along the spike, which  means that not only dipole radiation 
with $j=1$ but also higher modes travel along the spike and 
hence demonstrate that the spike remains effectively $3+1$ dimensional. This 
fact causes problem with a string interpretation of the spike 
soliton  because only $j=1$ mode should propagate on a fundamental 
string. A proposal to solve this puzzle was suggested in \cite{myers,myers1} 
where authors consider noncommutative solution describing 
N coincident D-strings attached to a D3-brane in nonabelian world-volume theory.
This solution accurately describes the physics 
at the core of the spike, very far from the D3-brane, and coincides
with the above spike soliton in abelian world-volume theory in the 
large N limit. The analysis shows that on noncommutative spike 
the mode spectrum is truncated at $l_{max} = N-1$ \cite{myers,myers1}.

In this article we are interested in studying physical 
effects which are produced on D3-brane 
by a spike which emanates normal to D3-brane world in the extra space direction.  
Our aim is to consider scattering of  electromagnetic 
plane waves on D3-brane spike using Dirac-Born-Infeld theory. 
Here one should take cake about the applicability of the results 
obtained in this effective theory and as we have seen 
much what this theory tell us is correct and we will take an 
uncritical attitude towards the 
validity of this approximation, to see how far we can get.
Specifically we are interested in 
studying reflected radiation produced by a spike
attached to our D3-brane world. We have observed that  
the spike sucks almost all electromagnetic radiation
and therefore acts like a black hole\footnote{The entire 
process in ten-dimensional $U(1)$ gauge theory is clearly 
unitary, but for the observer on the submanifold, on D3-brane,  
it is obviously not!}. 
This is because absorption cross section  for $j=1$ tends 
to a {\it constant} at low energy 
limit ($R\omega \rightarrow 0$) 
$$
\sigma^{abs}_1 ~ \sim ~ R^2 \rightarrow 
Const, ~~~~~\sigma^{abs}_j ~\sim R^2 ~(R\omega)^{4j-4} \rightarrow 0 ,~~j=2,3,...
$$

This behaviour is appealing for a string interpretation of the spike soliton 
because it is a constant for $j=1$ and tends to zero for higher $j > 1$,
therefore the propagation of $j=1$ mode is indeed distinctive 
\footnote{The apparent difference between this result and \cite{kastor} 
is that computing 
cross sections we have divided the scattered flux of radiation by 
the incident flux of electromagnetic field,
which is proportional to $\omega^2$ .}.
Instead, the scattered part of radiation on a D3-brane 
{\it tends to zero} for all $j=1,2,..$
$$
\sigma^{scat}_1 ~ \sim ~ R^2 (R\omega)^{2} ~\rightarrow ~0,~~~~~
\sigma^{scat}_j ~ \sim ~ R^2 (R\omega)^{8j-6} ~\rightarrow ~0,~~~~~ j=2,3,..
$$
demonstrating non-Thompson behaviour of the cross section $\sigma^{scat}_1$, 
universal for point like charged objects in electrodynamics. We formulate the 
comparison with the standard Thompson cross section in terms of the 
dynamical mass of the spike 
$$M^{2}_{dyn} = {e^4 \over R^2  (R\omega)^2 }.$$ 
This dynamical mass just describes the response of an $extended$ object 
to a perturbation of a wave length $\lambda = {1 \over \omega}$.
It is finite for nonzero $R\omega$, 
demonstrating that dynamical mass involved in the process  is 
finite and is of order of the wave length $1/\omega$. In the 
limit $\omega \rightarrow 0$ it tends to infinity demonstrating that 
static mass of the spike is infinite. Indeed our spike solution 
is infinitely long and has infinite mass by virtue having a constant 
energy per unit length. The small electromagnetic fluctuations on 
the static solution propagate with the $finite$ velocity of light
\cite{cm,konstantin}, therefore dynamical mass involved in the process is finite
and depends on the wave length of the perturbation. \footnote{The similar effect can be 
observed with the infinitely long strip of metal on one end of which acts mechanical 
perturbation. The crucial point is that the perturbation propagates 
with the finite velocity on an $extended$ object).}

{\it  The Lagrangian and the e\-qua\-ti\-ons. }
The nonlinear Dirac-Born-Infeld Lagrangian which contains both 
electric and magnetic fields, plus the scalar $x_9 \equiv \phi$ is
\cite{konstantin,bak}
\bea
L=-\int d^4x \sqrt{Det}~~~~~~~~~~~where~~~Det=1+ \vec{B}^2 - \vec{E}^2 -(\vec{E} \cdot \vec{B})^2 -
   (\partial_0 \phi)^2(1+\vec{B}^2) +  \nonumber\\  +(\vec{\partial}\phi)^2
   +(\vec{B}\cdot \vec{\partial} \phi)^2 - 
   - (\vec{E}\times \vec{\partial}\phi)^2+
   2\partial_0 \phi(\vec{B}[\vec{\partial} \phi\times\vec{E}]).~~~~~~~~~~~~\label{lagrfull}
\eea
and can also be written in a covariant form
$$
Det=1+ {1 \over 2} F_{\mu \nu} F^{\mu \nu} - \partial_{\mu} \phi \partial^{\mu} \phi
-({1 \over 8} \epsilon^{\mu \nu \lambda \rho} F_{\mu \nu} F_{\lambda \rho})^{2}
+({1 \over 2}\epsilon^{\mu \nu \lambda \rho} 
F_{\nu \lambda} \partial_{\rho} \phi)^{2}.
$$
The full set of field equations can be obtained by variation  
\bea
\partial_{\mu} \{  { F^{\mu \nu} (1 - \partial_{\rho} \phi \partial^{\rho}\phi) -
{1 \over 32}    ( F_{\lambda \rho} \tilde {F}^{\lambda \rho}) \tilde{F}^{\mu \nu} 
+ F^{\mu \rho} \partial_{\rho} \phi \partial^{\nu}\phi +
\partial^{\mu}\phi \partial_{\rho} \phi F^{\rho\nu} \over 
Det^{1/2} } \}= 0,
\label{coveq1}
\eea
\bea
\partial_{\mu} \{  { \partial^{\mu} \phi (1 - F_{\lambda \rho} F^{\lambda \rho}) +
2 F^{\mu \nu} F_{\nu \lambda} \partial^{\lambda} \phi \over Det^{1/2} } \}
= 0,~~~~~~~~~~~\epsilon^{\mu \nu \lambda \rho}\partial_{\nu} F_{\lambda \rho} =  0,
\label{coveq2}
\eea
where $\tilde{F}^{\mu \nu} = \epsilon^{\mu \nu \lambda \rho}F^{\lambda \rho} $.

We have to consider now small fluctuations around the background field 
configuration  of spike soliton (\ref{spike}) : 
$$ \vec{E}=\vec{E}_0 + \delta \vec{E},~~\vec{B}= 
\delta \vec{B},~~\phi=\phi_0 +\eta~.$$
Then keeping only terms in the $Det$  which are linear and quadratic
in the fluctuation one can get the resulting quadratic Lagrangian :
$2L_q = \delta\vec{E}^2(1+(\vec{\partial} \phi)^2) - \delta\vec{B}^2 
+ (\partial_0\eta)^2  -
(\vec{\partial}\eta)^2(1-\vec{E_0}^2) 
+ \vec{E_0}^2 
( \vec{\partial}\eta \cdot \delta\vec{E})~.$
Let us introduce the gauge potential for the fluctuation part of the e.m.
field as $(A_0,\vec{A})$ and substitute the values of the background 
fields from (\ref{spike})
\bea
2L_q = (\partial_0 \vec{A}- \vec{\partial}A_0)^2 (1+{R^4 \over r^4})-
       (\vec{\nabla}\times\vec{A})^2  \\ \nonumber
       + (\partial_0\eta)^2
       -(\vec{\partial}\eta)^2(1-{R^4 \over r^4}) + 
       {R^4 \over r^4} (\partial_0 \vec{A}- \vec{\partial}A_0)
       \cdot \vec{\partial}\eta ~, \label{fluctlag}
\eea
where we restore the dimensions of the fields and $R^2 = e^2 (2\pi \alpha^{'})$.
The equations that follow from this Lagrangian contain dynamical 
equations for the vector potential and for the scalar field, 
and a separate equation which represents a constraint. These
equations in the Lorenz gauge
$\vec{\partial}~\vec{A}=\partial_0 A_0$ are
\bea
 \label{alpha}
-\partial_0^2 \vec{A}(1+{R^4 \over r^4}) + \Delta\vec{A}+
{R^4 \over r^4}\vec{\partial}\partial_0(A_0+\eta) = 0 \\
\label{beta}
-\partial_0^2 A_0 + \Delta A_0 + 
\vec{\partial} {R^4 \over r^4} \vec{\partial}(A_0+\eta) - 
\vec{\partial} {R^4 \over r^4} \partial_0\vec{A} = 0  \\ 
\label{gamma}
-\partial_0^2 \eta~ + \Delta \eta~~ - 
\vec{\partial} {R^4 \over r^4} \vec{\partial}(A_0+\eta) + 
\vec{\partial} {R^4 \over r^4} \partial_0\vec{A} = 0
\eea
Equation (\ref{beta}) is a constraint: the time derivative of
the {\it lhs} is zero, as can be shown using the equation of motion 
(\ref{alpha}). One can choose $A_0=-\eta$ which is compatible 
with the field equations because the second and the third equations 
imply that $A_0 + \eta$ obeys the free wave equation.
With this condition the equations (\ref{beta}) and 
(\ref{gamma}) become the same, and
the first equation is also simplified \cite{konstantin}:
\be
\Delta\vec{A}~ - ~ (1+{R^4 \over r^4})~ \ddot{ \vec{A}}  = 0~,~~~~~~
\Delta \eta  ~ - ~ \ddot{\eta} ~ + ~ 
\vec{\partial} ({R^4 \over r^4} \dot{\vec{A}}) = 0~,
\label{seqns}
\qee
where $\dot{\vec{A}}$ denotes time derivative.
This should be understood to imply that once we obtain a solution,
$A_0$ is determined from $\eta$, but in addition we are now obliged to 
respect the gauge condition which goes over to 
$\vec{\partial}\vec{A}=-\dot{ \eta}$.

In the next section we shall consider the scattering of plane waves on 
a spike soliton (\ref{spike}). It is always possible to choose the 
plane wave to be a 1/2-BPS configuration. For that one should take  the 
plane wave solution of the  equations  (\ref{coveq1}),(\ref{coveq2}) in the form:                                              %
\be
A_{\mu} =e_{\mu} \phi_{0} e^{ikx},~~~~~~\phi = -\phi_{0} e^{ikx},
\qee
where $e^{\mu}_{\pm} = (\omega, \vec{e}_{\pm} +\vec{k})$,~ $\vec{e}_{\pm} *\vec{k} =0$,
and in addition we have  $\partial_{\mu} A^{\mu}=0,~~ A_0 + \phi =0$.

{\it Electromagnetic scattering on a spike soliton. }
Below we shall consider stationary scattering with definite energy 
(frequency $\omega$), therefore from (\ref{seqns}) we have 
\be
\Delta\vec{A}~ + ~ \omega^{2}~( 1+{R^4 \over r^4})~  \vec{A}  = 0,~~~~~ 
\Delta \eta  ~ + ~ \omega^{2}~ \eta ~- ~
\vec{\nabla}({i\omega R^4 \over r^4} \vec{A}) = 0~,
\label{seqnsf}
\qee
where $\vec{\nabla}\vec{A} = i\omega \eta$.
Let us consider following expansion for the vector potential :
$
\vec{A} = \vec{e} ~ \sum_{l} ~Y_{l0}(\theta) ~\zeta_{l} ,
$
where $\vec{e}$ is the constant polarization vector and $\zeta_{l}$
are the partial waves. From (\ref{seqns}) one can get equation which they should
satisfy 
\be
\partial^{2}_r  \zeta_{l} + {2 \over r} \partial_r \zeta_{l} 
+ [\omega^{2}( 1+{R^4 \over r^4}) - {l(l+1) \over r^2}]\zeta_{l}  = 0.
\qee
With the substitution similar to the one used in \cite{klebanov}
$
~~r = R e^{-z},~~\zeta_{l} = e^{-z/2}\tilde{\zeta}_{l}
$
it can be transformed into the form
\be
[\partial^{2}_z  + 2 (R\omega )^{2} ch 2z - (l +1/2)^2 ] \tilde{\zeta}_{l}=0. \label{transf}
\qee
This is the well known Mathieu's equation. The known mathematics of Mathieu's equation
\cite{dougall} allows to calculate a systematic expansion 
of solution in all orders in $(R\omega /2)$. The formulas 
needed for our purposes are summarized in \cite{gubser}. 
There are two asymptotic regions in our case:
the first one  $z     \gg  ln(R\omega )$  (I) where we have 
\be
[\partial^{2}_z  + (R\omega )^{2} e^{2z} - (l +1/2)^2 ] \tilde{\zeta}_{l}=0
\qee
and  the second  one $z \ll  - ln(R\omega )$ (II) where 
\be
[\partial^{2}_z  + (R\omega )^{2} e^{-2z} - (l +1/2)^2 ] \tilde{\zeta}_{l}=0.
\qee
The solutions in these two regions are known functions  
\be
\tilde{\zeta}^{I}_{l} = H^{(1)}_{l+1/2}(R\omega  e^{z}),~~~~~~~~
\tilde{\zeta}^{II}_{l} = A_{\infty}~ J_{l+1/2}(R\omega  e^{-z}) + 
B_{\infty}~N_{l+1/2}(R\omega  e^{-z}). \label{mainsolut}
\qee
The two regions overlap in the interval ~
$ ln(R\omega ) \ll  z \ll - ln(R\omega )$~ if ~$ R\omega \ll 1$.~
In this interval we can use the following expansion of the functions:
\bea
H^{(1)}_{l+1/2}(\xi)~ \simeq ~ (\xi/2)^{l+1/2}
{1 \over \Gamma(l+3/2)} ~+~ i(\xi/2)^{-l-1/2}{(-1)^{l+1} \over \Gamma(-l+1/2)} , \nonumber\\
J_{l+1/2}(\xi) ~ \simeq ~ (\xi/2)^{l+1/2}{1 \over \Gamma(l+3/2)},~~~~~
N_{l+1/2}(\xi) ~ \simeq ~ (\xi/2)^{-l-1/2}{(-1)^{l+1} \over \Gamma(-l+1/2)}  \nonumber\\
\nonumber
\eea
and thus to impose matching condition
\bea
({R\omega \over 2})^{l+1/2} {e^{(l+1/2)z} \over \Gamma(l+3/2)} + 
i(-1)^{l+1}({R\omega \over 2})^{-l-1/2} {e^{-(l+1/2)z}\over \Gamma(-l+1/2)} \simeq\\ \nonumber
A_{\infty}({R\omega \over 2})^{l+1/2} {e^{-(l+1/2)z} \over \Gamma(l+3/2)}
+B_{\infty} (-1)^{l+1}({R\omega \over 2})^{-l-1/2} {e^{(l+1/2)z}\over \Gamma(-l+1/2)}.
\eea 
From this one can find coefficients  $A_{\infty}$ and $B_{\infty}$
\bea
A_{\infty} = i(-1)^{l+1} ({R\omega \over 2})^{-2l-1} {\Gamma(l+3/2) \over \Gamma(-l+1/2)},~~~
B_{\infty} = (-1)^{l+1} ({R\omega \over 2})^{2l+1} {\Gamma(-l+1/2) \over \Gamma(l+3/2)}.
\eea
Using the asymptotic expansion for $H^{(1)}_{\nu}(\xi)$, $J_{\nu}(\xi)$ and $N_{\nu}(\xi)$
when  $\xi \rightarrow \infty$
and the formulas (\ref{transf}) we can recover the asymptotic behaviour of the 
solution at the origin where the spike is attached to the D3-brane 
\be
r\rightarrow ~0~~~~\zeta_{l}  \simeq ~~ \sqrt{ {2 \over \pi R\omega } } \label{origin}
e^{-i\pi(l+1)/2 } e^{{i\omega R^2 \over r}},
\qee
and at infinity from the origin
\bea   
r\rightarrow \infty~~~~\zeta_{l} \simeq ~~
{R \over  2r}~ \sqrt{ {2 \over \pi R\omega} }~ 
e^{i\pi(l+1)/2} ~( A_{\infty} + i B_{\infty}  ) [  e^{-i\omega r}  + 
R_{l}e^{i\omega r}]. ~~~~~~~~~~~~~\label{origininfty}
\eea
The first asymptotic (\ref{origin}) describes absorption waves which penetrate through the 
potential barrier  and propagate along the spike, the second one (\ref{origininfty}) 
describes  the scattered radiation. Thus the transmission and reflection amplitudes are equal to:
\be
T_l~~ =~~ {e^{-i \pi (l+1)/2}  \over {R \over 2}
e^{+i\pi(l+1)/2} ( A_{\infty} + i B_{\infty})  } ~~=~~
-i({2 \over R})~ ({R\omega \over 2})^{2l+1} ~ {\Gamma(-l+1/2) \over 
\Gamma(l+3/2)}, \label{tran31}
\qee
\be
R_l = {e^{-i \pi (l+1)/2} ( A_{\infty} - i B_{\infty})  \over 
e^{+i\pi(l+1)/2} ( A_{\infty} + i B_{\infty})  } ~  =  ~                   
(-1)^{l+1}[ 1 - 2  ({R\omega \over 2})^{4l+2} ~ ({\Gamma(-l+1/2) \over 
\Gamma(l+3/2)})^2].\label{reflec33}
\qee
For the lower partial waves they are:
\be
\begin{array}{lcl}
T_0 = -2i\omega ,&~~~&R_0 = -1 + 2 (R \omega )^2.
\end{array}
\qee
and for the higher partial waves  they behave as:
$
T_l \sim \omega (R\omega)^{2l},~~~~~~R_l - (-1)^{l+1} \sim (R\omega)^{4l+2}.
$

{\it Cross Sections.}
The asymptotic behaviour of the wave function can be represented
in the form:
\bea
\vec{A} = \vec{e}~~ (e^{ikz} + {f(\theta) \over r} 
e^{ikr} + g(\theta)~ e^{ikR^{2}/r}~)= ~~~~~~~~~~~~~~~~~~~~~~~~~~~~\label{asympt}\\ \nonumber
=\vec{e}~ \sum^{\infty}_{l=0} ~{\sqrt{4\pi (2l+1)} \over 2ik (-1)^{l+1} }~ Y_{l0}(\theta)~
[{1\over r}e^{-ikr}  + (-1)^{l+1}{1\over r}(1 + 2ik  f_{l})~ e^{ikr} + 
(-1)^{l+1}~2ik  g_{l}~ e^{ikR^{2}/r} ],    
\eea
where 
$
f(\theta) = \sum^{\infty}_{l=0} ~\sqrt{4\pi (2l+1)}~ Y_{l0}(\theta)~ f_l 
$
and comparing this with the asymptotic solution (\ref{origin}), (\ref{origininfty}) 
and   (\ref{tran31}), (\ref{reflec33})   we can find that 
$R_l = (-1)^{l+1}(1 +2i\omega f_l) $ and $T_{l} = (-1)^{l+1} 2i\omega g_l$ and 
therefore:
\be
f_{l} =  {i \over \omega}  ({R\omega \over 2})^{4l+2}({\Gamma(-l+1/2) \over \Gamma(l+3/2)})^2 ,~~~
g_{l} = (-1)^{l} {1 \over R \omega}({R\omega \over 2})^{2l+1} ~ {\Gamma(-l+1/2) \over \Gamma(l+3/2)}.
\qee
In particular for low $l$ they are equal to
$
f_0 = i R (R\omega),~f_1 = {i \over 9} R (R\omega)^5
,~g_0 = 1,~g_1 = {1 \over 3}  (R\omega)^2 .
$

If $dI$ is the energy radiated by the system into spherical angle $d\Omega$ and
incident electromagnetic plane wave has Pointing vector $\vec{S}$  then the
cross section is equal to
$
d\sigma = {dI \over S}.
$ 
The Pointing vector $\vec{S}$ for the incident plane wave $ \vec{e}~ e^{ikz}$ is equal to
$
\vec{S}_{in} = {1 \over 4\pi} B^{2}_{in}~ \vec{e}_z =  {\omega^{2} \over 4\pi}~ \vec{e}_z.
$
For the scattered radiation $ \vec{e}~{f(\theta) \over r} e^{ikr}$ and absorption 
$ \vec{e}~g(\theta)  e^{ikR^{2}/r}$ waves it is 
$
\vec{S}_{scat} = ~ \vec{n}{1 \over 4\pi} B^{2}_{scat},~~~~~\vec{S}_{abs} = 
- \vec{n}{1 \over 4\pi} B^{2}_{abs},~~ \vec{n} 
= {\vec{k} \over k}.
$
Thus we have to find $\vec{B}_{scat}$ and $\vec{B}_{abs}$, ~
$
\vec{B}_{scat}= \vec{\nabla}\times \vec{A} = i[\vec{k} \times \vec{e}] ~{f(\theta) \over r} e^{ikr},~~~
\vec{B}_{abs}= -i[\vec{k} \times \vec{e}] ~g(\theta)~ {R^{2} \over r^2} e^{ikR^{2}/r},
$
and then the  
$
\vec{S}_{scat} =~ \vec{n}{\omega^{2} \over 4\pi}~\vert  
\vec{n} \times \vec{e} \vert^{2}~ {\vert f(\theta) \vert^2  \over r^2},$ and 
$\vec{S}_{abs} =-~ \vec{n}{\omega^{2} \over 4\pi}~\vert  
\vec{n} \times \vec{e} \vert^{2}~ \vert g(\theta) \vert^2  {R^4 \over r^4}.
$
The radiated energy into the spherical angle $d\Omega$ thus will be
\be
dI_{scat} = {\omega^{2} \over 4\pi}~\vert  
\vec{n} \times \vec{e} \vert^{2}~ \vert f(\theta) \vert^2 ~d\Omega,~~~~ 
dI_{abs} = {\omega^{2} \over 4\pi}~\vert  
\vec{n} \times \vec{e} \vert^{2}~ \vert g(\theta) \vert^2 {R^4 \over r^2} ~d\Omega.
\qee
Finally for the cross sections  we have expressions:
\be
{d\sigma_{scat} \over d\Omega}  = \vert \vec{n} \times \vec{e} 
\vert^{2}~ \vert f(\theta) \vert^2,~~~~~~{d\sigma_{abs} \over d\Omega^{'}} = 
R^{2}~ \vert \vec{n} \times \vec{e} \vert^{2}~ \vert g(\theta) \vert^2 . \label{cross}
\qee
In the last formula for the absorption cross section we used the symmetry of the system 
$(r^{'} \rightarrow R^2/r)$ mentioned in \cite{thesis,kastor} and replaced the spherical angle 
$d\Omega (R^2/r^{2})$ into $d\Omega^{'}$ simply because $ dr^{'} \rightarrow -(R/r)^{2}dr$

We can also find the corresponding cross sections for the scalar field
$\phi$ using the fact that $i\omega \phi = \vec{\nabla}\vec{A}$. Thus
we have:
\be
{d\sigma^{\phi}_{scat} \over d\Omega}  = \vert \vec{n} \cdot \vec{e} 
\vert^{2}~ \vert f(\theta) \vert^2,~~~~~~{d\sigma^{\phi}_{abs} \over d\Omega^{'}} = 
R^{2}~ \vert \vec{n} \cdot \vec{e} \vert^{2}~ \vert g(\theta) \vert^2 . 
\label{scalcross}
\qee
The total cross section of electromagnetic and scalar fields is the sum 
$\sigma + \sigma^{\phi} $. As it is easy to see the total absorption and scattering
cross sections do not depend on polarization vector \cite{konstantin} because 
$
\vert \vec{n} \times \vec{e} \vert^{2} + \vert \vec{n} \cdot \vec{e} \vert^{2} =1.
$
To evaluate the integral we have to calculate the vector product 
$
\vert \vec{n} \times \vec{e} \vert^{2} = cos^{2}\theta cos^{2}\varphi + sin^{2}\varphi,
$
therefore for the total cross section we get
\be
\sigma = {1 \over 2} \int (1+ cos^{2}\theta) \vert f(\theta) \vert^2 ~d\Omega
\qee
Using the formulas
$
cos^{2}\theta = {4\pi \over 3} Y_{10} ~ Y^{*}_{10}
$
and expanding the product 
$\sqrt{{4\pi \over 3}}~ Y_{10} *Y_{lm}$ over $ Y_{l+1m}$ and $Y_{l-1m}$
one can evaluate completely the integration in (\ref{cross}) and get expression for the 
cross section
\be
\sigma_{scat} = 4\pi \sum^{\infty}_{l=0}  {(3l^2 +3l -2) \over (2l-1)(2l+3) } ~(2l+1)~ \vert f_l \vert^2 ~
+~2\pi \sum^{\infty}_{l=0}  {(l +1)(l+2) \over (2l+3) } ~ (f_l f^{*}_{l+2} + f_{l+2}f^{*}_l )
\qee 
Using the expression for the amplitudes $f_l$ we can find the lowest term in the 
scattering cross section 
\be
\sigma^{scat}_1 = {8\pi \over 3} \vert f_0 \vert^2 =  
{8\pi \over 3} R^2  (R\omega)^2. \label{radi}
\qee
We have to compare this cross section with the Thompson cross section and for 
that reason we shall rewrite our result in the Thompson from:
$$
\sigma_{T} = {8\pi \over 3} {e^4 \over M^{2}_{dyn}}
$$
where $M_{dyn}$ is the frequency dependent dynamical mass of the 
extended charged object. The behaviour
of these cross section (\ref{radi}) as a function of $\omega$ is essentially 
different and comes as a big surprise, because the
cross section instead of approaching 
a constant tends to zero! The only way to explain this result is to  
analyze the dynamical mass of the extended spike 
\be
M^{2}_{dyn} = {e^4 \over R^2  (R\omega)^2 }.
\qee
We observe that it is finite for nonzero $R\omega$, demonstrating 
that dynamical mass involved in the process  is 
finite and is of order of the wave length $1/\omega$. In the low energy 
limit it tends to infinity demonstrating that "static" mass of the spike is infinite. 
Indeed the spike soliton is infinitely long and has infinite mass by virtue having a constant 
energy per unit length. The small electromagnetic fluctuations on 
this static solution propagate with the $finite$ velocity of light
\cite{cm,konstantin}, therefore dynamical mass involved in the process is finite
and depends on the wave length of the perturbation. 

In the same way one can find that the absorption cross section 
\be
\sigma^{abs}_1 = {8\pi \over 3}R^{2} \vert g_0 \vert^2 =  {8\pi \over 3} R^2, 
\qee
and contrary to the scattering one (\ref{radi}) tends to a constant! Thus we see that 
the spike soliton sucks almost all 
electromagnetic radiation and therefore acts like a black hole.
This result corresponds to the dipole radiation when we consider electromagnetic
perturbation propagating on the spike in the form of P-wave  (j=1) 
which is coming down the spike with a
polarization along a direction parallel to the brane \cite{konstantin}. 

We can also compute the corresponding cross sections for the scalar field,
the only difference is in the coefficient: instead of $8\pi/3$ it is  $4\pi/3$ 
and the total cross section $\sigma + \sigma^{\phi}$ comes with  coefficient $4\pi$.

{\it Partial electromagnetic  waves.}
In previous section we used partial wave expansion, but it does 
not correspond to the physical expansion over total angular momentum 
of the  electromagnetic field. To do so and to distinguish corresponding
partial waves with definite parity we have to expand the vector potential 
over the vector spherical harmonics:
$
\vec{Y}^{el}_{jm} = {1 \over \sqrt{j(j+1)}} \nabla_{\vec{n}}  
Y_{jm},~~P=(-1)^{j};~
\vec{Y}^{mag}_{jm} = [\vec{n} \times 
\vec{Y^{el}_{jm}}],~~P=(-1)^{j+1};~
\vec{Y}^{lon}_{jm}  = \vec{n} Y_{jm},~~P=(-1)^{j},
$
where $j=1,2,3,...$ and the corresponding parity is shown. For 
any given angular momentum  $j$ there is one even and one odd 
parity state. The  fields carrying angular momentum $j$ and the 
parity $(-1)^{j}$ are electric multipoles $Ej$ and the parity 
$(-1)^{j+1}$ are magnetic multipoles $Mj$. 

Using this basic functions we can rederive  the expansion (\ref{asympt}) in          %
the form
\be
\vec{A}_{scat}= {1 \over 2}\sum_{j=1,\lambda=0\pm1}^{\infty}~ \sqrt{4\pi 
(2l+1)}~ (\vec{e}*\vec{e}^{~*}_{\lambda})~
\{ f^{el}_{j\lambda}~ \vec{Y}^{el}_{j\lambda}+
f^{mag}_{j\lambda} ~\vec{Y}^{mag}_{j\lambda} + 
f^{lon}_{j\lambda}~ \vec{Y}^{lon}_{j\lambda} \} {e^{ikr} \over r}
\qee
where $\vec{e}_0 = \vec{e}_z$,~$\vec{e}_{\pm 1} = \vec{e}_x \pm \vec{e}_y$ 
($\vec{e}\cdot\vec{e}_0 =0 $) and 
\be
\begin{array}{lrlr}
f^{el}_{j\lambda} = -\lambda~{jf_{j+1} +(j+1)f_{j-1} \over 2j+1},&\lambda=\pm 1;&f^{el}_{j\lambda}=
{\sqrt{j(j+1)} \over 2j+1}(-f_{j+1} + f_{j-1}),&\lambda=0;\\
f^{mag}_{j\lambda} = ~if_{j},&\lambda=\pm 1;&f^{m}_{j\lambda} =0,&\lambda=0;\\
f^{lon}_{j\lambda} = -\lambda {\sqrt{j(j+1)} 
\over 2j+1}(-f_{j+1} + f_{j-1}),&\lambda=\pm 1;&f^{lon}_{j\lambda}
={(j+1)f_{j+1} +jf_{j-1}  \over 2j+1},&\lambda=0.
\end{array}
\qee
In the last formula the amplitudes $f^{el}_{j\lambda}, f^{mag}_{j\lambda}$  
($\lambda = \pm 1$) describe physical transverse 
degrees of freedom of electromagnetic field  and using 
the fact that $i\omega \phi = \vec{\nabla}\vec{A}$ 
we can identify the amplitude $f^{lon}$ as describing scattering of the scalar field.

The partial cross sections can be represented now in the form:
\be
\sigma^{scat}_{j} = 2\pi (2j+1) \vert f^{(\delta)}_{j} \vert^2 ,
\qee
where index $\delta =el,~mag$ describes electric or magnetic multipoles. For the 
electric dipole, quadruple and higher multipoles the amplitudes are:
$
f^{el}_1 = {2f_0 + f_2 \over 3},~~f^{el}_2 = 
{3f_1 + 2f_3 \over 5},~~f^{el}_j \simeq {j+1 \over 2j+1}f_{j-1}
$
and the corresponding scattering cross sections are:
\be
\sigma^{el}_1 =  6 \pi \vert f^{el}_1 \vert^2,~~~
\sigma^{el}_2 =  10\pi \vert f^{el}_2 \vert^2,~~~~~~\sigma^{el}_j = {\pi \over 2}
{(j+1)^2 \over (2j+1)} ({\Gamma(-j+3/2) \over \Gamma(j+1/2)})^4 R^2 (R\omega/2)^{8j-6} .
\qee 
For the magnetic dipole and quadruple radiation we have 
\be
\sigma^{mag}_1 =  6 \pi  \vert f_{1} \vert^2,~~~
\sigma^{mag}_2 =  10 \pi \vert f_{2} \vert^2 ,~~~\sigma^{mag}_j = {\pi(2j+1) \over 2}
({\Gamma(-j+1/2) \over \Gamma(j+3/2)})^4 R^2 (R\omega/2)^{8j+2}.
\qee
Similar formulas are valid for the absorption amplitude $\vec{e}~g(\theta)  e^{ikR^{2}/r}$,
therefore for electric multipoles we have:
\be
\sigma^{abs}_j = {\pi \over 2}
{(j+1)^2 \over (2j+1)} ({\Gamma(-j+3/2) \over \Gamma(j+1/2)})^2 R^2 (R\omega/2)^{4j-4}
\qee 
and one can also get less dominant magnetic multipoles.

The general asymptotic behaviour of the partial scattering cross sections have the form:
\be
\sigma^{scat}_j \sim R^2 (R\omega)^{8j-6},~~~~~ j=1,2,3,..
\qee
This behaviour essentially differs from the one in electrodynamics 
being $\omega^{2j -2}$.
The absorption cross section is:
\be
\sigma^{abs}_1 \sim R^2 \rightarrow Const, ~~~~~~\sigma^{abs}_j \sim R^2 (R\omega)^{4j-4},~~j=2,3,...
\qee
this behaviour is appealing for the string interpretation of the spike soliton 
because it is a constant for $j=1$ and tends to zero for higher $j > 1$,
therefore the propagation of $j=1$ mode is indeed distinctive.

We are grateful to H.Nielsen and P.Olesen for stimulating discussions. 
One of the authors (G.S.) is indebted to the Leipzig University for kind hospitality.
This work was supported in part by the EEC Grant no. HPMF-CT-1999-00162.

{\it Appendix. }
In previous sections the scattering and absorption amplitudes have been calculated
by solving equation (\ref{transf}) in lowest order in $\lambda = \frac{1}{2} R \omega$ by
standard continuity argument. The known mathematics of Mathieu's equation
\cite{dougall} allows to calculate systematic expansion in all
orders in $\lambda$. The formulae needed for our purpose have been
summarized in \cite{gubser} (see also \cite{manvelyan}). 
The transition $(T_{l})$ and reflection $(R_{l})$ partial amplitudes
are (instead of equations (\ref{tran31}), (\ref{reflec33}))

\begin{eqnarray}
T_{\ell} = -2i \chi_0^{-1} { \frac{1}{2} [ 1+(1+ \sum_1^{\infty} \lambda^{4k}
\overline \mu_k)^{-2}] (1 + \sum_1^{\infty} \lambda^{4k} \chi_k)^{-1} \over
[ 1 + \chi_0^{-2} (1+\sum_1^{\infty} \lambda^{4k} \chi_k)^{-2}  
 (1+\sum_1^{\infty} \lambda^{4k} \overline \mu_k)^{-2} ] }  , \cr
 \\
R_{\ell} = (-1)^{\ell} {  [ 1- \chi_0^{-2} (1+ \sum_1^{\infty} \lambda^{4k}
\chi_k)^{-2}] (1 + \sum_1^{\infty} \lambda^{4k} \overline \mu_k)^{-1} \over
[ 1 + \chi_0^{-2} (1+\sum_1^{\infty} \lambda^{4k} \chi_k)^{-2}  
 (1+\sum_1^{\infty} \lambda^{4k} \overline \mu_k)^{-2} ] } .
\end{eqnarray}

We have evaluated the first correction ${\cal O}(\lambda^4
)$ only:
$
\chi_0  = \lambda^{-(2 \ell +1)} \ \ {\Gamma(\ell + \frac{3}{2}) \over
\Gamma(\frac{1}{2} - \ell ) }, \ \  \ \mu_0 = \frac{1}{2} (\ell + \frac{1}{2}
) \ \
\mu_1 = {1 \over \ell + \frac{1}{2} } \ ( {1 \over \ell + \frac{3}{2} } - {
1\over \ell - \frac{1}{2} }) , \    
\chi_1 =2 \mu_1 \ [  \psi (\ell + \frac{3}{2} ) + \psi(\frac{1}{2} - \ell) - 2
\ln \lambda ] 
- {1 \over (\ell + \frac{3}{2} )^2 } - { 1 \over (\ell - \frac{1}{2} )^2}.
$
The first order correction changes the leading small $R \omega$ behaviour
of the scattering amplitude
$
R_{\ell } = 1 - 8 \lambda^2   \delta_{\ell, 0} + \lambda^4 (2 \pi i \mu_1(\ell )
 + 32 \ \delta_{\ell, 0} ) +
{\cal O}(\lambda^8, \lambda^{4\ell +2}) .             
$
This implies that all partial scattering cross sections for $j \geq 1$
behave like
\begin{equation} \sigma_{j}^{scatt} \sim R^2 (R\omega)^6 
\end{equation}
with a coefficient decreasing at large $j$ like $j^{-5 }$.
This is an essential change compared to the one anticipated from the zero
approximation, $ \sigma_{j}^{scatt} \sim R^2 (R\omega)^{8j -6}$.

The corrections to the transition amplitude are less significant
$
T_{\ell} = 2 \lambda^{2\ell +1} {\Gamma(\frac{1}{2} -\ell) \over 
\Gamma(\ell + \frac{3}{2}) } 
(1 - 4 \lambda^2 \delta_{\ell,0} - \lambda^4 \ (\ 2 \pi i \mu_1(\ell)  +
 \chi_1(\ell) ) 
+ {\cal O}(\lambda^8, \lambda^{4\ell+2}).
$
The leading behaviour for $R\omega \rightarrow 0$ is really the one expected
from zero approximation.


\end{document}